\documentstyle[preprint,aps,prb]{revtex}
\begin{document}
\title{Spin correlation functions \\ in random-exchange $s=1/2$ $XXZ$ chains}
\author{Heinrich R\"oder$^a$, Joachim Stolze$^a$, Richard N. Silver$^b$,
and Gerhard M\"uller$^c$}
\address{$^a$Physikalisches Institut, Universit{\"a}t Bayreuth, 95440
Bayreuth, Germany\\
$^b$Theoretical Division, LANL, Los Alamos, New Mexico 87545\\
$^c$Department of Physics, The University of Rhode Island, Kingston,
Rhode Island 02881-0817}

\date{\today}
\maketitle

\begin{abstract}
The decay of (disorder-averaged) static spin correlation functions at
$T=0$ for the one-dimensional spin-1/2 $XXZ$ antiferromagnet with uniform
longitudinal coupling $J\Delta$ and random transverse coupling $J\lambda_i$
is investigated by numerical calculations for ensembles of finite chains.
At $\Delta=0$ ($XX$ model) the calculation is based on the Jordan-Wigner
mapping to free lattice fermions for chains with up to $N=100$ sites.
At $\Delta \neq 0$ Lanczos diagonalizations are carried out for chains
with up to $N=22$ sites.
The longitudinal correlation function $\langle S_0^z S_r^z\rangle$
is found to exhibit a power-law decay with an exponent that varies with
$\Delta$ and, for nonzero $\Delta$, also with the width of the
$\lambda_i$-distribution.
The results for the transverse correlation function
$\langle S_0^x S_r^x\rangle$ show a crossover from power-law decay to
exponential decay as the exchange disorder is turned on.
\end{abstract}


The combination of randomness and quantum fluctuations is well-known to be
a fertile ground for interesting physical phenomena including
Anderson localization.
In one-dimensional (1D) tight-binding systems the rule is that disorder
always leads to localization.
However, if the randomness is purely off-diagonal, the localization length
diverges at the band center,\cite{ER78} which is bound to affect the decay
law of correlation functions.
One particular off-diagonally disordered fermion model, the 1D
half-filled tight-binding model with random hopping, is equivalent to
the special $XX$ case $(\Delta=0)$ of the 1D $s=1/2$ $XXZ$ model with
random transverse exchange coupling, described by the Hamiltonian
\begin{equation}
H= J \sum_i [\lambda_i (S_i^x S_{i+1}^x +
S_i^y S_{i+1}^y) + \Delta S_i^z S_{i+1}^z] \;.
\label{1}
\end{equation}
The uniform longitudinal spin coupling corresponds to a fermion interaction.
Here we consider the range $0 \leq \Delta \leq 1$, use periodic boundary
conditions, and take the random transverse coupling $J \lambda_i$ to be
described by a Gaussian distribution with
$\overline{\lambda_i}=1$, $\overline{\lambda_i\lambda_j}$ =
$1+\sigma^2\delta_{ij}$.
The spin correlations at $T=0$ of this model were recently investigated
by means of a real-space renormalization group (RSRG) method\cite{DF92,F94}
based in part on ideas from earlier work,\cite{DM80} and by means of a
finite-chain study.\cite{HRD93}

One interesting proposition made in the context of the RSRG
study is the existence of a {\em random-singlet phase}
with algebraically decaying spin pair correlations:\cite{DF92,F94}
\begin{equation}
\langle S_0^\alpha S_r^\alpha\rangle \sim (-1)^r r^{-\eta_\alpha}, \;\;
\alpha = x, z \;.
\label{2}
\end{equation}
The singlet nature of that phase would imply that the characteristic
exponent $\eta_\alpha$ assumes the same value in the longitudinal $(z)$
and transverse $(x)$ correlation functions, in marked contrast to the
case with no exchange disorder $(\sigma=0)$, for which we know the
exact result\cite{LP75}
\begin{equation}
\eta_x = 1/\eta_z = 1 - (1/\pi)\arccos\Delta \;.
\label{3}
\end{equation}
The RSRG study further predicts that this exponent value is
$\eta_x = \eta_z =2$, independent of the longitudinal coupling $\Delta$ and
the disorder strength $\sigma$, provided the latter is not too small.
It is indeed quite unusual that the anisotropic randomization of an
anisotropic exchange interaction should effectively remove the effects of
anisotropy in the spin correlations.

Here we report results of a finite-chain study which goes significantly
beyond that of Ref. \onlinecite{HRD93} in statistics and system sizes.
For the $XX$ model $(\Delta=0)$, we carry out the computation in the (free-)
fermion representation, which enables us to handle chains with up to
$N=100$ spins and beyond.
For $\Delta \neq 0$ we must resort to Lanczos diagonalizations. Here the
largest system for which we can perform the computation with reasonable
statistics has $N=22$ sites.
For graphical purposes we shall consider, henceforth, the absolute value,
$|\langle S_0^\alpha S_r^\alpha\rangle|$, of the spin pair
correlations.

We first consider the case $\Delta=0$ ($XX$ model).
If the longitudinal correlation function does exhibit power-law decay,
$|\langle S_0^z S_r^z\rangle|$ $\sim$ $r^{-\eta_z}$, as predicted,
then the exponent $\eta_z$ also governs the $N$-dependence of the function
$|\langle S_0^z S_{N/2}^z\rangle|$ in a cyclic chain of $N$
sites.\cite{HRD93}
We have evaluated this quantity for systems with $N\leq 100$ sites and
for disorder strengths $\sigma\leq 2$, all with ensemble averages
over up to $10^5$ configurations.

The data analysis yields $\eta_z=2$ independent of $\sigma$.
This is consistent with the RSRG prediction\cite{DF92,F94} but in
contradiction to the earlier finite-size study,\cite{HRD93} where a
significant $\sigma$-dependence of $\eta_z$ was observed.\cite{note1}
Our data also confirm that the disorder-averaged logarithm of
$|\langle S_0^z S_r^z\rangle|$ exhibits the decay law
$\sim \exp(-r^{1/2})$ as predicted in Ref. \onlinecite{F94}.

The decay of the transverse correlation function
$\langle S_0^x S_r^x\rangle$ is much
more sensitive to the presence of exchange disorder, as is demonstrated
by the data shown in Figs. \ref{F1} and \ref{F2}.
In the main diagram of Fig. \ref{F1} we show the function
$|\langle S_0^x S_r^x\rangle|$ versus $r$ in a logarithmic plot for
ensembles with different disorder strengths.
Turning on the exchange disorder with gradually increasing
$\sigma$ causes the transverse correlations to decay more and more
rapidly as one might expect.\cite{note2}

For $0 \leq \sigma \lesssim 0.5$ the data describe a power-law behavior.
This is also evident in the bundle of curves near the top of Fig. \ref{F2},
which shows the $r$-dependence of the function
$|\langle S_0^x S_r^x\rangle|$ semi-logarithmically at $\sigma=0.4$ for
various system sizes.
The values $|\langle S_0^x S_{N/2}^x\rangle|$ at the endpoints of these
curves plotted vs $N/2$ in a log-log graph fall onto a straight line,
and the slope of that line determines the exponent $\eta_x$.
This is illustrated by the full squares in the inset to Fig. \ref{F2}.

The $\sigma$-dependence of $\eta_x$ as obtained from this procedure is
shown in the inset to Fig. \ref{F1}.
For the system without exchange disorder we reproduce the exactly known
value $\eta_x=1/2$,\cite{M68} which is a special case of (\ref{3}).
As $\sigma$ increases from zero, $\eta_x$ grows gradually and
monotonically, at first slowly, then more and more rapidly.

For $\sigma \gtrsim 0.5$ the curves in Fig. \ref{F1} suggest the
occurrence of a crossover from algebraic decay to exponential decay,
$|\langle S_0^x S_r^x\rangle|$ $\sim$ $\exp(-r/\xi)$, in the range of
$r$ for which we have data.
The exponential character of the decay is more strikingly manifest in
the lower bundle of data shown in the main plot of Fig. \ref{F2},
representing the function $|\langle S_0^x S_r^x\rangle|$ at $\sigma=1$
for various system sizes.\cite{note2}
The smallest expectation values are known only with considerable (relative)
uncertainty despite the augmented statistics.

The triangles, which represent the values
$|\langle S_0^x S^x_{N/2}\rangle|$ vs $N/2$ in this semi-logarithmic plot,
are consistent with a straight line.
Its slope determines the disorder-induced correlation length $\xi$.
Over the range of disorder strengths, where our data suggest exponential
decay of $|\langle S_0^x S_r^x\rangle|$, $\xi$ thus determined decreases
monotonically with increasing $\sigma$.

Our data for the exchange disordered $XX$ model are consistent with two
alternative scenarios, which are equally interesting:
(i) There exists a transition at some nonzero value of the disorder
strength, $\sigma_c \simeq 0.5$, from algebraically to exponentially
decaying transverse spin correlations.
(ii) A transition of the same nature occurs at $\sigma_c = 0$ instead,
which produces very similar crossover effects in the finite-chain data.
A more extensive study for longer chains and with better statistics
will be necessary to discriminate with confidence between the two
scenarios.\cite{prel}
The data are definitely incompatible with a persistent power-law decay
as predicted by RSRG.

Now we turn to one case, $\Delta=0.75$, with fermion interaction
($XXZ$ model).
Since the computations are much more involved, the available data are
limited by comparison with the case $\Delta=0$.
At $\Delta \neq 0$ neither our data for the longitudinal correlations nor
those for the transverse correlations are compatible with the RSRG
predictions.

The function $|\langle S_0^z S_r^z \rangle|$ for various system sizes and
$\sigma=1.5$ is shown logarithmically in Fig. \ref{F3}.
The endpoint data $(r=N/2)$, which fall neatly onto a straight line,
describe a power-law decay with $\eta_z = 1.26$.
The exponent values obtained for two smaller disorder strengths are
$\eta_z = 1.17$ $(\sigma=0.5)$ and $\eta_z = 1.31$ $(\sigma=0.25)$.
The exact result (\ref{3}) for $\sigma=0$ assumes the value
$\eta_z = 1.298\ldots$

All combined, the data suggest that the function
$|\langle S_0^z S_r^z \rangle|$ is governed by a power-law which persists
in the presence of randomness.
The $\sigma$-dependence of the exponent $\eta_z$ appears to go through a
minimum of considerable depth at $\sigma \neq 0$,\cite{note3}
which implies the curious phenomenon that the longitudinal correlations
are enhanced by a small amount of transverse exchange disorder relative to
the correlations in the uniform-exchange system.

The data for the transverse correlations $|\langle S_0^x S_r^x \rangle|$
at $\Delta=0.75$ exhibit properties very similar to what we have observed
and described for the free-fermion case ($\Delta=0$).
For not too large disorder strengths $(\sigma \lesssim 0.5)$, we see a
power-law behavior with an exponent that increases monotonically from
the exactly known value $\eta_x = 0.769\ldots$ at $\sigma=0$, as given
by expression (\ref{3}), to $\eta_x = 1.00$ at $\sigma=0.25$ and
$\eta_x = 1.49$ at $\sigma=0.5$, at which point a crossover to exponential
behavior makes itself felt.
The exponential decay law at $\sigma=1.5$ is quite evident in the
semi-logarithmic plot of Fig. \ref{F4}.

The discrepancies between our results and the RSRG predictions of Refs.
\onlinecite{DF92,F94} call for an explanation in future studies.
Possibly, the strongly {\em anisotropic} nature of the exchange
in the model system (\ref{1}) -- even for $\Delta=1$ -- is not adequately
taken into account by the RSRG procedure, which derives from a method
originally developed for a model with {\em isotropic} exchange.\cite{DM80}

In order to gain further insight into the properties of the $XXZ$ chain
with random exchange, we plan to investigate the nature of low-lying
excitations and the properties of dynamic correlation functions.
At $\Delta=0$ the Jordan-Wigner mapping to free fermions will make it
possible to carry out these calculations for large systems.
At $\Delta \neq 0$ the KPM method developed recently\cite{SR94} promises
to be an adequate calculational instrument.

The work at URI was supported by NSF Grant DMR-93-12252 and by the NCSA at
Urbana-Champaign.


\begin{figure}
\caption[one]
{Log-log plot of $|\langle S_0^x S_r^x \rangle|$ at $T=0$ in the
random-exchange $XX$ chain $(\Delta=0)$ with $N=40$ spins at disorder
strengths $\sigma = 0, 0.1, \ldots, 1$ (top to bottom).
For $0<\sigma\leq 0.5$ the expectation value of
$\langle S_0^x S_r^x \rangle$ has been averaged over $10^4$ configurations,
and for $\sigma > 0.5$, over $10^5$ configurations. The inset shows the
(effective) decay exponent $\eta_x$ as a function of the disorder strength
$\sigma$.}
\label{F1}
\end{figure}

\begin{figure}
\caption[two]
{Semi-logarithmic plot of $|\langle S_0^x S_r^x \rangle|$ with $r\leq N/2$
at $T=0$ in the random-exchange $XX$ chain $(\Delta=0)$ with $N$ = 18, 22,
26, 30, 34, 40, for $\sigma=0.4$ (upper set of curves, averaged over $10^4$
configurations) and $\sigma=1$ (lower set of curves, averaged over $10^5$
configurations).
The data points for $\sigma=1$ at maximum distance $(r = N/2)$ are marked
by full triangles.
The straight line which best fits these data points is shown dot-dashed
and determines the correlation length $\xi$.
The inset shows the data for $\sigma=0.4$ in a log-log plot.
The data points at maximum distance $(r=N/2)$ are marked by full squares.
The straight line which best fits these data points is shown
dot-dashed and determines the correlation exponent $\eta_x$.}
\label{F2}
\end{figure}

\begin{figure}
\caption[three]
{Log-log plot of $|\langle S_0^z S_r^z \rangle|$ for the $XXZ$ model with
uniform longitudinal exchange $(\Delta=0.75)$ and random transverse
exchange $(\sigma=1.5)$ on chains of various lengths.
The data points represent averages over 1000 configurations for
$N=6,\ldots,18$, and 450 configurations for $N=22$.}
\label{F3}
\end{figure}

\begin{figure}
\caption[four]
{Semi-logarithmic plot of $|\langle S_0^x S_r^x \rangle|$ for the $XXZ$
model with uniform longitudinal exchange $(\Delta=0.75)$ and random
transverse exchange $(\sigma=1.5)$ on chains of various lengths.
The data points represent averages over 1000 configurations for
$N=6,\ldots,18$, and 450 configurations for $N=22$.}
\label{F4}
\end{figure}



\begin{references}
\bibitem{note2} The upward curvature in the vicinity of $r=N/2$ is an
obvious finite-size effect attributable to periodic boundary conditions.
\bibitem{prel} Preliminary results for chains with up to $N=200$ sites
indicate that scenario (ii) is more likely.
\bibitem{note3} The same effect had previously been observed in Ref.
\onlinecite{HRD93}, also for the case $\Delta=0$, where it is absent
in our data.
\end{references}
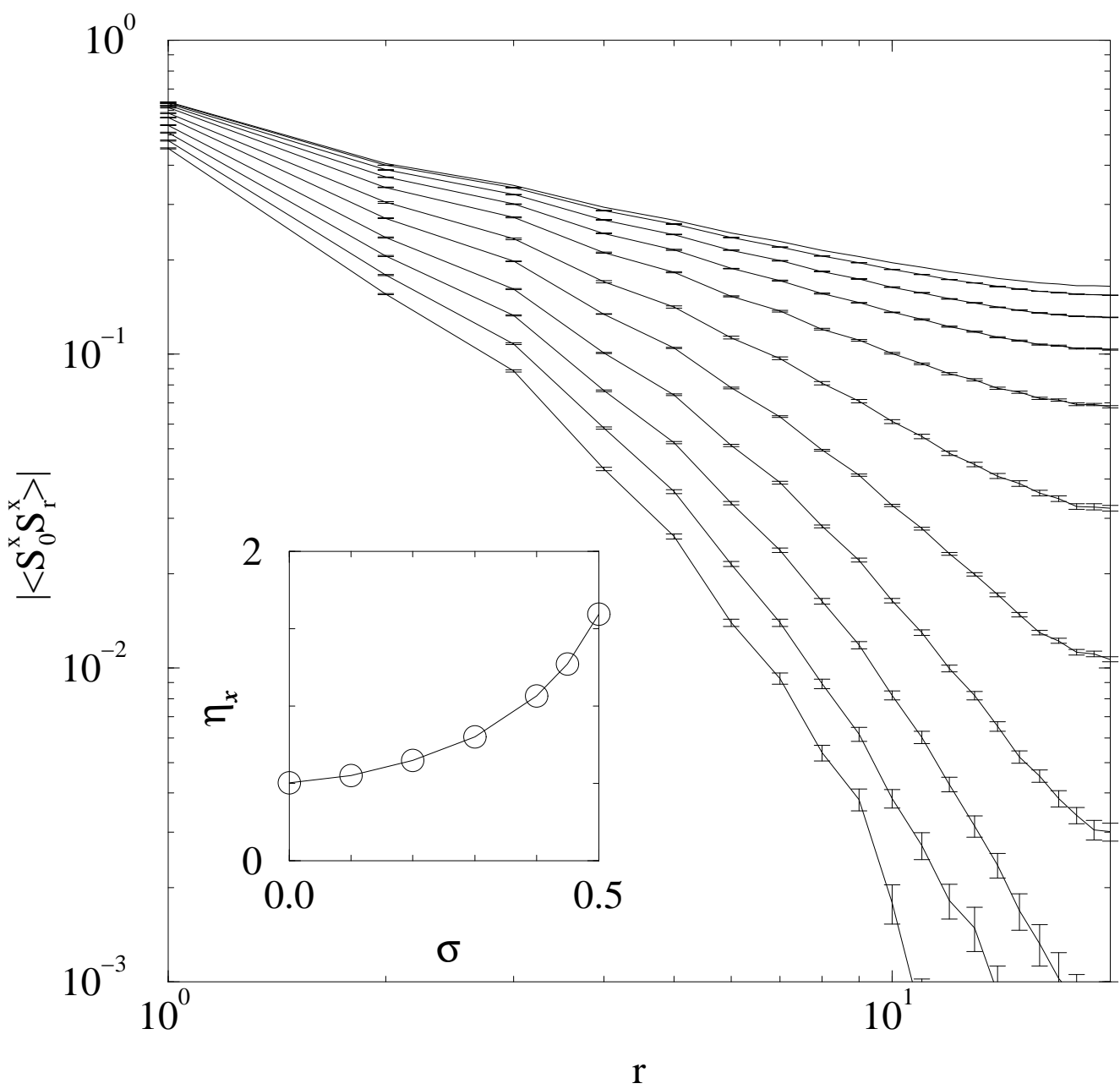
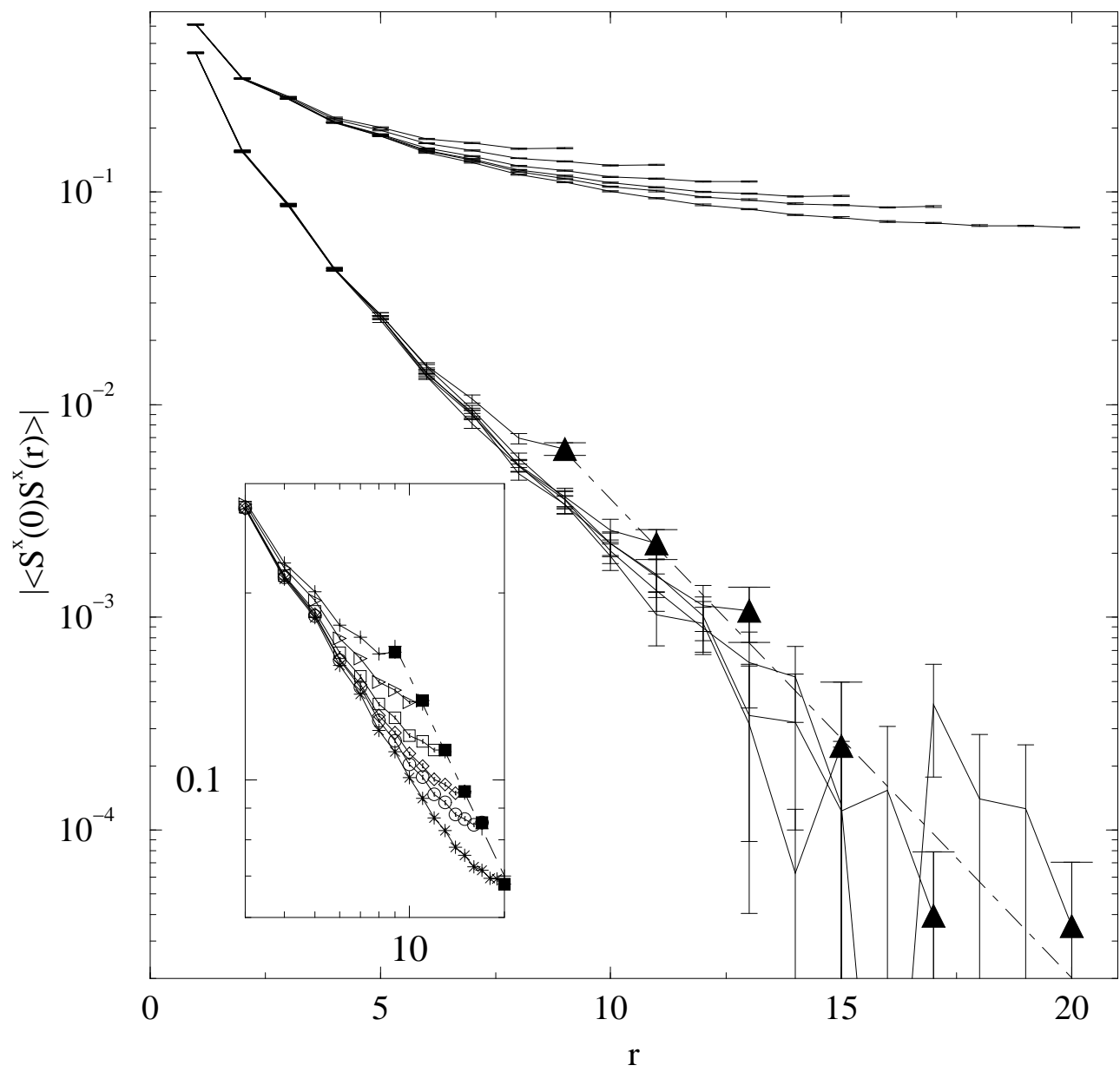
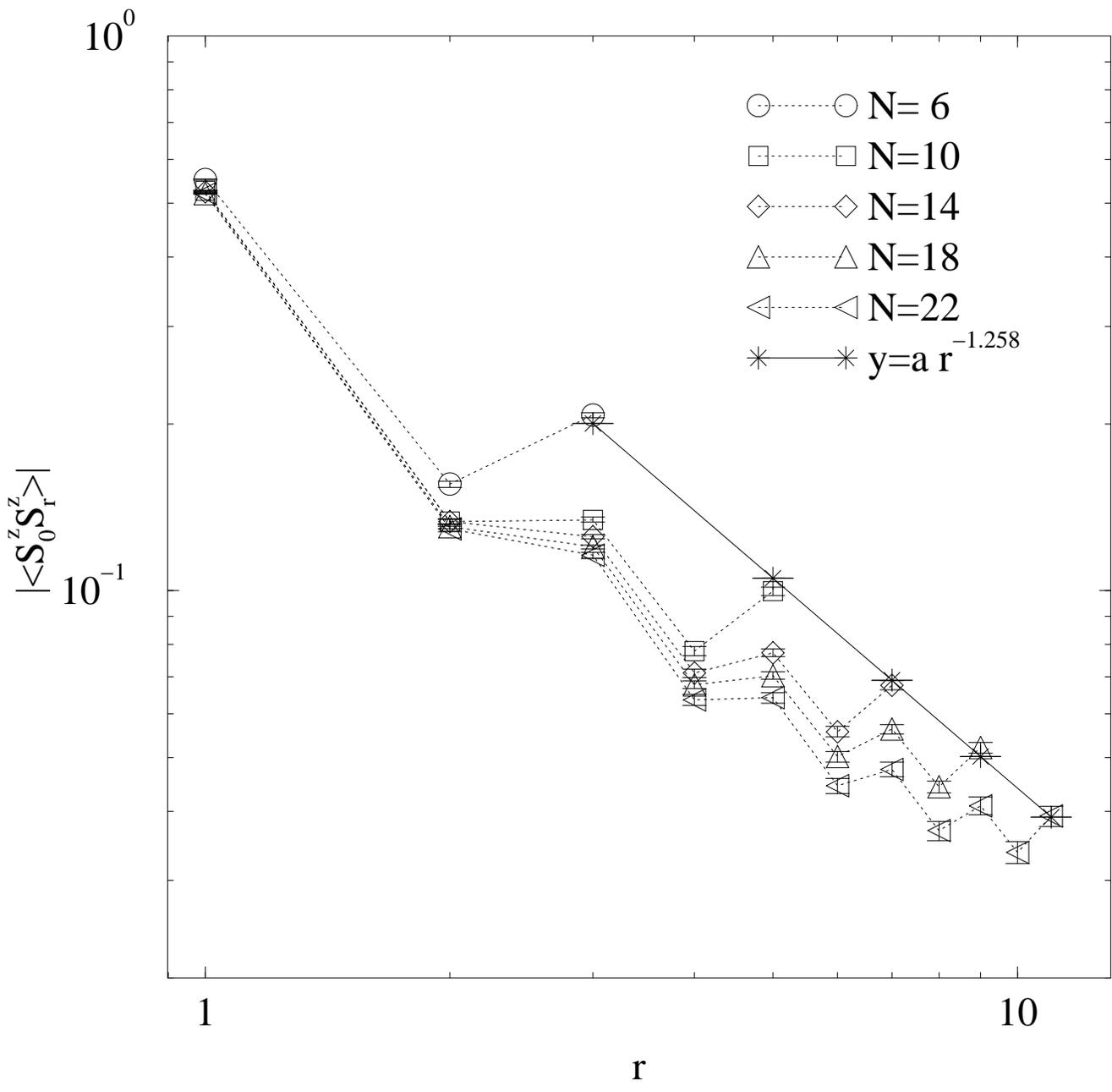
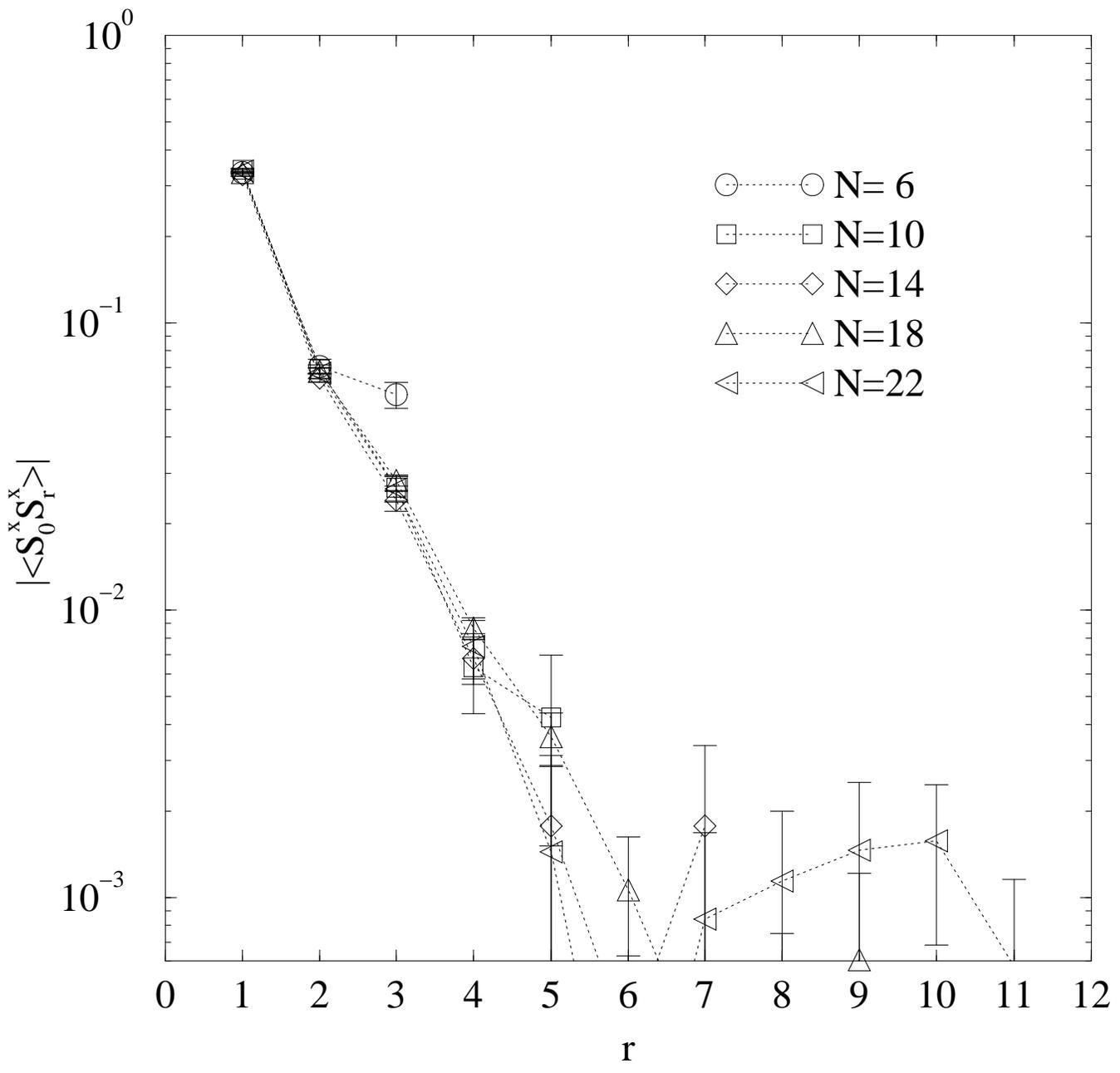
\end{document}